\documentclass[aps,prx,10pt,twocolumn,superscriptaddress]{revtex4}

\usepackage{graphicx}
\usepackage{amssymb, amsmath}
\usepackage{color}
\usepackage{ulem}
\usepackage{bm}
\usepackage{graphicx}
\usepackage{subfigure}
\usepackage{dcolumn}
\usepackage{mathtools}
\def\I{\uppercase\expandafter{\romannumeral 1}}
\def\II{\uppercase\expandafter{\romannumeral 2}}
\def\III{{\uppercase\expandafter{\romannumeral 3}}}
\def\IV{{\uppercase\expandafter{\romannumeral 4}}}
\def\V{{\uppercase\expandafter{\romannumeral 5}}}
\def\VI{{\uppercase\expandafter{\romannumeral 6}}}
\def\VII{{\uppercase\expandafter{\romannumeral 7}}}
\def\i{\lowercase\expandafter{\romannumeral 1}}
\def\ii{\lowercase\expandafter{\romannumeral 2}}
\def\iii{{\lowercase\expandafter{\romannumeral 3}}}
\def\iv{{\lowercase\expandafter{\romannumeral 4}}}
\def\v{{\lowercase\expandafter{\romannumeral 5}}}
\def\vi{{\lowercase\expandafter{\romannumeral 6}}}
\def\vii{{\lowercase\expandafter{\romannumeral 7}}}

\def\angstrom{\mbox{\normalfont\AA}}

\def\nn{\nonumber\\}

\def\angstrom{\mbox{\normalfont\AA}}
\def\k{\mathbf{k}}
\def\nn{\nonumber\\}

\begin{document}

\title{Quantum valley Hall effect, orbital magnetism, and  anomalous Hall effect in twisted multilayer graphene systems }

\author{Jianpeng Liu}
\affiliation{Department of Physics, Hong Kong University of Science and Technology, Kowloon, Hong Kong}

\author{Zhen Ma}
\affiliation{School of Physics, Huazhong University of Science and Technology, Wuhan 430074, China}

\author{Jinhua Gao}
\affiliation{School of Physics, Huazhong University of Science and Technology, Wuhan 430074, China}

\author{Xi Dai}
\affiliation{Department of Physics, Hong Kong University of Science and Technology, Kowloon, Hong Kong}

\begin{abstract}
We study the electronic structures and topological properties of $(M+N)$-layer twisted graphene systems. We consider the generic situation that $N$-layer graphene is placed on top of the other $M$-layer graphene, and is twisted with respect to each other by an angle $\theta$. In such twisted multilayer graphene (TMG) systems, we find that there exists two low-energy flat bands for each valley emerging from the interface between the $M$ layers and the $N$ layers. These two low-energy bands in the TMG system possess valley Chern numbers that are dependent on both the number of layers and the stacking chiralities. In particular, when the stacking chiralities of the $M$ layers and $N$ layers are opposite,  the total Chern number of the two low-energy bands for each valley equals to $\pm(M+N-2)$ (per spin). If the stacking chiralities of the $M$ layers and the $N$ layers are the same, then the total Chern number of the two low-energy bands for each valley is  $\pm(M-N)$ (per spin).
 The valley Chern numbers of the low-energy bands are associated with large, valley-contrasting orbital magnetizations, suggesting the possible existence of orbital ferromagnetism and anomalous Hall effect once the valley degeneracy is lifted either externally by a weak magnetic field or internally by Coulomb interaction through spontaneous symmetry breaking.

\end{abstract}

\maketitle

Twisted bilayer graphene (TBG) has drawn significant attention recently due to the observations of the correlated insulating phases \cite{cao-nature18-mott,sharpe-tbg-19,choi-tbg-stm,kerelsky-tbg-stm,marc-tbg-19} and unconventional superconductivity \cite{cao-nature18-supercond,marc-tbg-19}.
At small twist angles, the low-energy states of TBG are characterized by four low-energy bands contributed by the two nearly decoupled monolayer valleys \cite{santos-tbg-prl07,macdonald-pnas11}. Around the ``magic angles", the bandwidths of the four low-energy bands become vanishingly small, and these nearly flat bands are believed to be responsible for most of those exotic properties observed in TBG. Numerous theories have been proposed to understand the electronic structures \cite{po-prx18, yuan-prb18, koshino-prx18, kang-prx18, song-tbg-18, po-tbg2, origin-magic-angle-prl19, pal-kindermann-arxiv18, liu-ll-arxiv, ll-tbg-lian,zhang-tbg-arxiv19}, the correlated insulating phase \cite{po-prx18,sboychakov-arxiv-18, isobe-prx18, xu-lee-prb18, huang-arxiv-18,liu-prl18,rademaker-prb18, venderbos-prb18, kang-tbg-correlation-arxiv18, xie-tbg-2018, jian-moire,zaletel-tbg-2019}, and the mechanism of superconductivity \cite{xu-prl18,po-prx18,isobe-prx18,wu-prl18,wu-xu-arxiv-18,lian-arxiv-18, huang-tbg-19,liu-prl18, venderbos-prb18, nematic-tbg-arxiv18, wu-tbg-chiral-arxiv18, roy-tbg-prb19}. 

On the other hand, interesting topological features have already emerged in the electronic structure of TBG.
It has been shown that the four low-energy bands are topologically nontrivial in the sense that they are characterized by odd windings of Wilson loops \cite{song-tbg-18,yang-tbg18,liu-ll-arxiv}, which is an example of the fragile  topology \cite{po-tbg2}. The four flat bands have been further proposed to be equivalent to the zeroth pseudo Landau levels (LLs) with opposite Chern numbers and sublattice polarizations \cite{liu-ll-arxiv}, which is the origin of the nontrivial band topology in the TBG system.  

Moreover, recently unconventional ferromagnetic superconductivity and correlated insulating phase have been observed in twisted double bilayer graphene \cite{kim-double-bilayer-arxiv19,cao-double-bilayer-arxiv19}. It implies that the low-energy flat bands, which are believed to be responsible for the correlated physics in TBG, may also exist in the twisted double bilayer graphene system. A recent theoretical study indeed revealed the presence of flat bands in twisted double bilayer graphene \cite{chebrolu-arxiv19}. 
Motivated by these works, in this paper  we study the electronic structures and topological properties of twisted multilayer graphene (TMG). In particular, we consider the most generic situation that the $N$-layer chirally stacked graphene is placed on top of the other $M$-layer chirally stacked graphene, and they are twisted with respect to each other by a non-vanishing angle $\theta$, as schematically shown in Fig.~\ref{fig1}(a) (for the case of $M\!=\!2, N\!=\!2$). In such a ($M+N$)-layer TMG system, we propose that there always exists two low-energy bands (for each valley), and that the bandwidths  of the two low-energy bands become vanishingly small at the magic angles of twisted bilayer graphene (TBG)  for arbitrary numbers of layers $M$ and $N$. The flat bands in the TMG system can be interpreted from the pseudo LL representation of TBG \cite{liu-ll-arxiv}, and is protected by an approximate chiral symmetry in chiral graphene multilayers. 

Moreover, we also find that there is a  Chern-number hierarchy in the $(M+N)$-layer TMG system. In particular, when the stacking chiralities of the $M$ layers and $N$ layers are the same,  the total Chern number of the two low-energy flat bands for each monolayer valley equals to $\pm(M-N)$ for each spin species 
\footnote{Unless otherwise specified, in the rest of the paper the Chern number refers to the Chern number for each physical spin speicies.}. On the other hand, if the stacking chiralities of the $M$ layers and the $N$ layers are opposite, then the total Chern number of the two low-energy bands for each valley is  $\pm(M+N-2)$. The valley Chern numbers can be further tuned by an external electric field, leading to gate-tunable quantum valley Hall effect.

The nonzero valley Chern numbers of the low-energy flat bands are characterized by large and valley-contrasting orbital magnetizations.
With the presence of an external magnetic field or the spontaneous symmetry breaking induced by the Coulomb interactions, the valley degeneracy is expected to be broken, and a valley-polarized (quantum) anomalous Hall state may be realized. The valley polarized state are associated with chiral current loops, which generate local magnetic fields peaked at the $AA$ region. The local magnetic fields generated by the chiral current loops may be a robust experimental signature for the nonzero valley Chern number and the valley polarized state in the TMG system.
The flat bands at  the universal magic angles, together with the Chern-number hierarchy and orbital magnetism, make the TMG systems a unique platform to study  strongly correlated physics with nontrivial band topology, and may have significant implications on the observed ferromagnetic superconductivity and correlated insulating phase in twisted double bilayer graphene \cite{kim-double-bilayer-arxiv19,cao-double-bilayer-arxiv19}.

\section{Electronic structures of the twisted multilayer graphene systems}

\subsection{The lattice structures}
\begin{figure}
\includegraphics[width=3.0in]{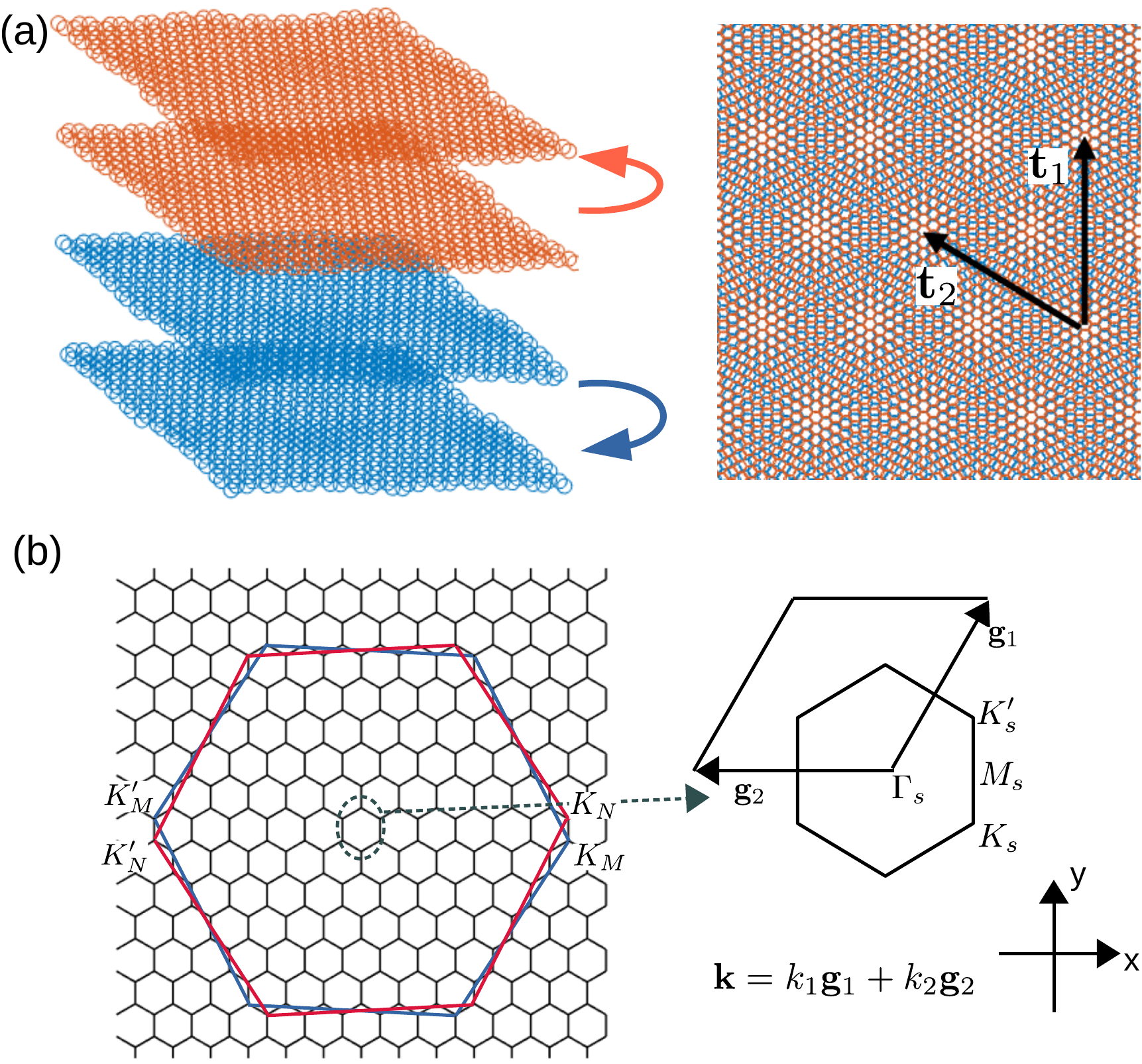}
\caption{(a) Left: structure of the twisted multilayer graphene with $M\!=2\!$ and $N\!=\!2$ (see text).   Right: moir\'e pattern of the twisted multilayer graphene, seen from the top. $\mathbf{t}_1$ and $\mathbf{t}_2$ denoted the lattice vectors of the moir\'e supercell. (b) The Brillouin zones of the top $N$ multiplayers, bottom  $M$ multilayers, and the moire\'e supercell are plotted in red, blue, and black lines respectively.}
\label{fig1}
\end{figure}
We consider the most generic case of chirally stacked twisted multilayer graphene, i.e., we place $N$ chiral graphene multilayers on top of $M$ chiral graphene multilayers, and twist them with respect to each other by an angle $\theta$. This is schematically shown in Fig.~\ref{fig1}(a) for the case of $M\!=\!2,N=2\!$. Similar to the case of TBG, commensurate moir\'e supercells are formed when the twist angle $\theta(m)$ obeys the condition 
$\cos{\theta(m)}=(3m^2+3m+1/2)/(3m^2+3m+1)$ \cite{castro-neto-prb12}, where $m$ is a positive integer. The  lattice vectors of the moir\'e superlattice are expressed as $\mathbf{t}_1=(-\sqrt{3}L_s/2,L_s/2)$, and $\mathbf{t}_2=(0,L_s)$, where $L_s=\vert\mathbf{t}_1\vert=a/(2\sin{(\theta/2)})$ is  the size of the moir\'e supercell, and $a=2.46\,$\angstrom\ is the lattice constant of graphene.  In TBG it is well known that there are atomic corrugations, i.e., the variation of interlayer distances on the moir\'e length scale. In particular, in the $AB (BA)$ region of TBG,  the interlayer distance $d_{AB}\!\approx\!3.35\,$\angstrom\, while in the $AA$-stacked region the interlayer distance
$d_{AA}\!\approx\!3.6\,$\angstrom\ \cite{graphite-AA}. Such atomic corrugations  may be modeled as \cite{koshino-prx18}
\begin{equation}
d_z(\mathbf{r})=d_{0}+2d_1\sum_{j=1}^{3}\cos{(\,\mathbf{g}_j\!\cdot\!\mathbf{r}\,)}\;,
\label{eq:dz-1}
\end{equation}
where $\mathbf{g}_1$, $\mathbf{g}_2$, and $\mathbf{g}_3\!=\!\mathbf{g}_1+\mathbf{g}_2$ are the three reciprocal lattice vectors of the moire supercell. We take $d_0=3.433\,$\angstrom\ and $d_1=0.0278\,$\angstrom\ in order to reproduce the interlayer distances in $AA$- and $AB$-stacked bilayer graphene. In this paper, the atomic corrugations of the two twisted layers at the interface (between the $M$ layers and the $N$ layers) is also be modeled by Eq.~(\ref{eq:dz-1}). On the other hand, the interlayer distances within the untwisted $M$ layers and the untwisted $N$ layers are set to the interlayer distance of Bernal bilayer graphene $d_{AB}\!=\!3.35\,$\angstrom. At a small twist angle $\theta$, the Brillouin zone (BZ) of the moir\'e supercell has been significantly reduced  compared with those of the untwisted multilayers as shown in Fig.~\ref{fig1}(b). 

\subsection{The effective Hamiltonian}

The low-energy effective Hamiltonian of the twisted $(M+N)$-layer TMG of the $K$ valley is expressed as 
\begin{equation}
H^{K}_{\alpha,\alpha'}(M+N)=\begin{pmatrix}
H^{K}_{\alpha}(M) & \mathbb{U} \\
\mathbb{U}^{\dagger} & H^{K}_{\alpha'}(N)
\end{pmatrix}\;,
\label{eq:HMN}
\end{equation}
where $H^{K}_{\alpha}(M)$ and $H^{+}_{\alpha'}(N)$ are the effective Hamiltonians for the $M$-layer and $N$-layer graphene with stacking chiralities $\alpha, \alpha'=+/-$. In particular, 
\begin{equation}
H^{K}_{\alpha}(M)=\begin{pmatrix}
h_{0}(\k) & h_{\alpha} & 0 & 0 & ... \\
h_{\alpha}^{\dagger} & h_{0}(\k) & h_{\alpha} & 0 & ...\\
0 & h_{\alpha}^{\dagger} & h_{0}(\k) & h_{\alpha} & ...\\
 & & & ... &   
\end{pmatrix}\;,
\label{eq:HM}
\end{equation}
where  $h_{0}(\k)\!=\!-\hbar v_{F}(\k-\mathbf{K}_M)\cdot\mathbf{\sigma}$ stands for the low-energy effective Hamiltonian for monolayer graphene near the Dirac point $\mathbf{K}_M$, and $h_{\alpha}$ is the interlayer hopping, with
\begin{equation}
h_{+}=\begin{pmatrix}
0 & 0\;\\
t_{\perp} & 0 
\end{pmatrix}\;,
\label{eq:chiral-hopping}
\end{equation}
and $h_{-}=h_{+}^{\dagger}$.

The off-diagonal term $\mathbb{U}$ represents the coupling between the twisted $M$ layers and $N$ layers. Here we assume that there is only the nearest neighbor interlayer coupling, i.e., the topmost layer of the $M$-layer graphene is only coupled with the bottom-most layer of the $N$-layer graphene, thus   
\begin{equation}
\mathbb{U}=\begin{pmatrix}
0 & ... & 0 \\
\vdotswithin{0} & ... & 0 \\
U(\mathbf{r})e^{-i\Delta\mathbf{K}\cdot\mathbf{r}} & ... & 0
\end{pmatrix}\;,
\label{eq:twist-coupling1}
\end{equation}
where the $2\!\times\!2$ matrix $U$ describes the tunneling between the Dirac states of the twisted bilayers \cite{macdonald-pnas11,koshino-prx18}
\begin{equation}
U(\mathbf{r})=\begin{pmatrix}
u_0 g(\mathbf{r}) & u_0'g(\mathbf{r}-\mathbf{r}_{AB})\\
u_0'g(\mathbf{r}+\mathbf{r}_{AB}) & u_0 g(\mathbf{r})
\end{pmatrix}\;,
\label{eq:u}
\end{equation}
where $\mathbf{r}_{AB}\!=\!(\sqrt{3}L_s/3,0)$,  $u_0'$ and $u_0$ denote the intersublattice and intrasublattice interlayer tunneling amplitudes, with $u_0'\!\approx\!0.098\,$eV, and $u_0\!\approx\!0.078$\,eV \cite{koshino-prx18}. $u_0\!$ is  smaller than $u_0'$ due to the effects of atomic corrugations \cite{koshino-prx18, liu-ll-arxiv}. $\Delta\mathbf{K}=\mathbf{K}_N-\mathbf{K}_M=(0,4\pi/3L_s)$ is the shift between the Dirac points of the $N$ layers and the $M$ layers. The phase factor $g(\mathbf{r})$ is defined as $g(\mathbf{r})=\sum_{j=1}^{3}e^{i\mathbf{q}_j\cdot\mathbf{r}}$, with $\mathbf{q}_1=(0,4\pi/3L_s)$, $\mathbf{q}_2=(-2\pi/\sqrt{3}L_s,-2\pi/3L_s)$, and $\mathbf{q}_3=(2\pi/\sqrt{3}L_s,-2\pi/3L_s)$. It worth to note that Eq.~(\ref{eq:HMN}) is the effective Hamiltonian for the $K$ valley.
The Hamiltonian for the $K'$ valley is readily obtained by applying a time-reversal operation to $H^{K}_{\alpha,\alpha'}(M+N)$. 

\subsection{The emergence of two flat bands and the universal magic angles}
\begin{figure}
\includegraphics[width=3.5in]{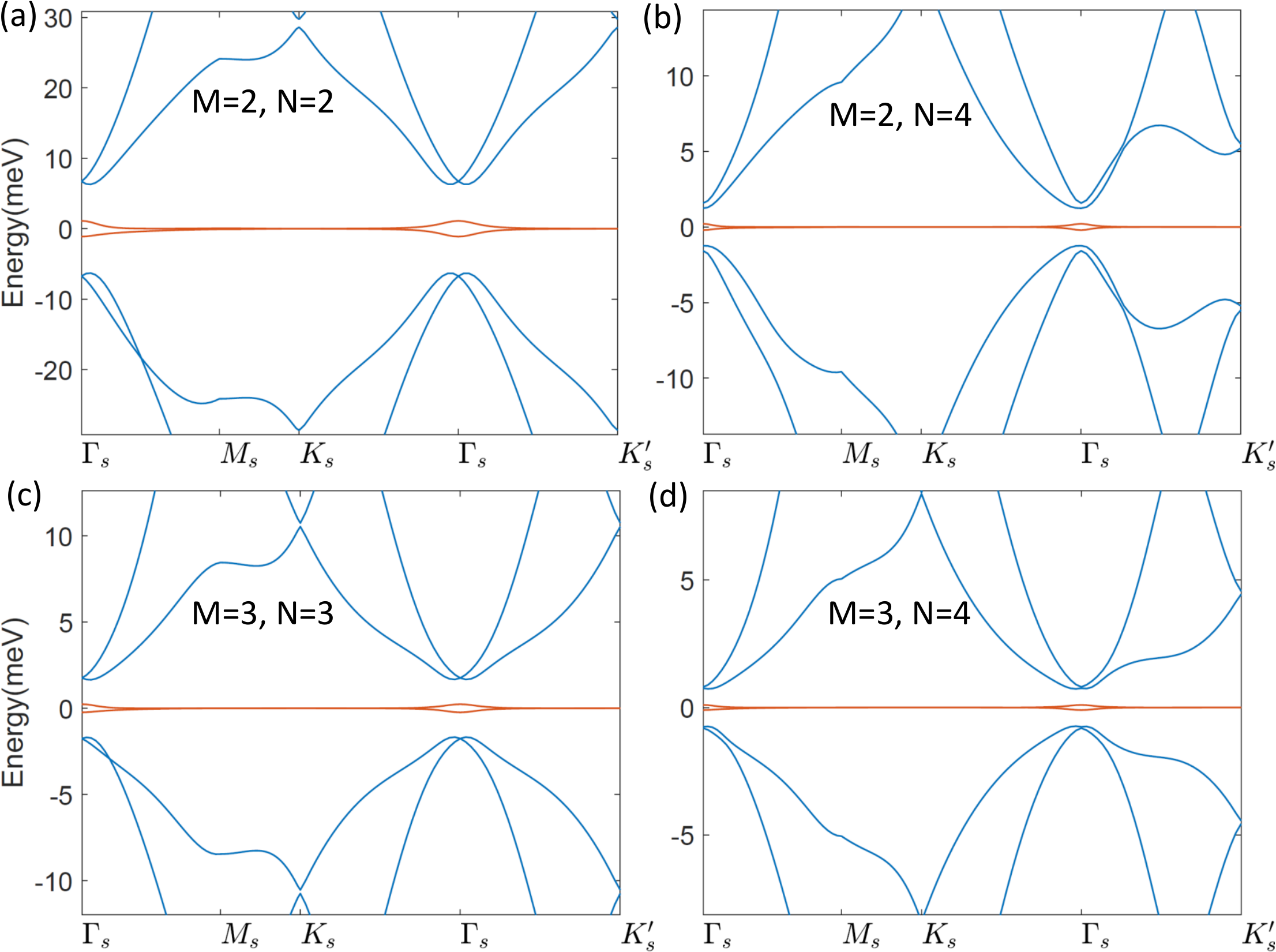}
\caption{The bandstructures of $(M+N)$-layer twisted multilayer graphene at the first magic angle $\theta\!=\!1.05^{\circ}$: (a)$M\!=\!2, N\!=\!2$, (b)$M\!=\!2, N\!=\!4$, (c) $M\!=\!3, N\!=\!3$, and (d) $M\!=\!3, N\!=\!4$. }
\label{fig2}
\end{figure}
We continue to study the electronic structures of the $(M+N)$-layer TMG systems using the effective Hamiltonian given by Eq.~(\ref{eq:HMN}).  The bandstructures for ($M\!=\!2$, $N\!=\!2$), ($M\!=\!2$, $N\!=\!4$), $(M\!=\!3, N\!=\!3)$, and $(M\!=\!3,N\!=\!4)$ at the first magic angle of TBG $\theta\!=\!1.05^{\circ}$ with the same stacking chiralities ($\alpha\!=\!\alpha'=+$) are shown in Fig.~\ref{fig2}(a)-(d) respectively \footnote{The bandstructures with opposite stacking chiralities are very similar to those shown in Fig.~\ref{fig2}.}. Clearly there are two low-energy flat bands marked by the red lines that are separated from the other bands. The two low-energy bands are almost exactly flat at $\theta\!=\!1.05^{\circ}$ for all these TMG systems with different layers, indicating that the magic angle of TBG is universal for the TMG systems regardless the number of layers. It turns out that the two flat bands in TMG originates from the twisted bilayer at the interface, and they remain flat even after being coupled with the other graphene layers due to an (approximate) chiral symmetry of Eq.~(\ref{eq:HMN}). More details about the origin of the flat bands in the TMG systems can be found in Appendix ~\ref{sec:append2}.   

\begin{figure}
\includegraphics[width=3.5in]{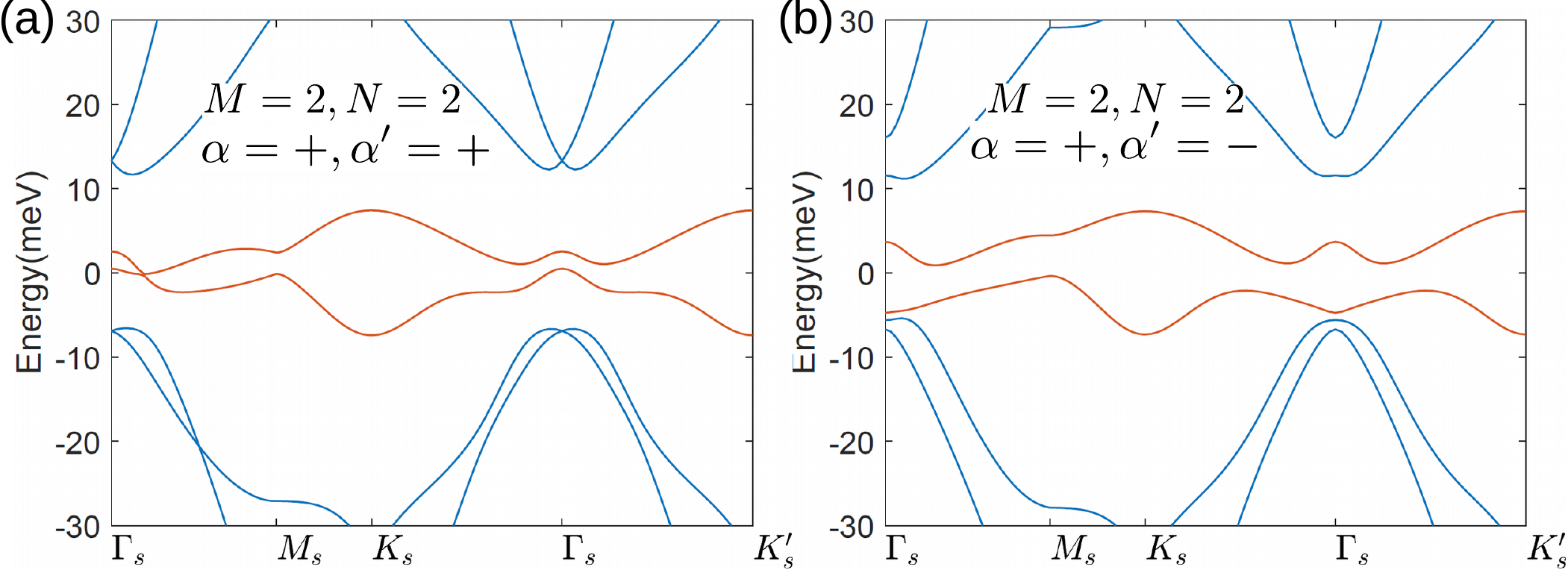}
\caption{The bandstructures of $(2+2)$-layer twisted multilayer graphene at the first magic angle $\theta\!=\!1.05^{\circ}$ with the more realistic interlayer hopping Eq.~(\ref{eq:chiral-hopping-new}): (a)the two bilayers have the same stackign chirality, and (b) the two bilayers have the opposite stacking chiralities. }
\label{fig2b}
\end{figure}

In realistic situations there are also further neighbor interlayer hoppings in graphene multilayers, which would break the chiral symmetry of the effective Hamiltonian in Eq.~(\ref{eq:HMN}), and the flat bands shown in Fig.~\ref{fig2} would become more dispersive. In order to test the robustness of the flat bands, we have included all the second-neighbor and third-neighbor interlayer hoppings with intersite distances equal to $\sqrt{a^2/3+d_{AB}^2} $ and $\sqrt{a^2+d_{AB}^2}$ respectively ($d_{AB}\!\approx\!3.35\,$\angstrom is the interlayer distance), and their amplitudes are denoted by $t_2$ and $t_3$. After including these terms, the interlayer hopping term with $+$ stacking chirality becomes 
\begin{equation}
h_{+}=\begin{pmatrix}
t_2f(\k) & t_2 f^*(\k)\;\\
t_{\perp}-3t_3 & t_2 f(\k) 
\end{pmatrix}\;,
\label{eq:chiral-hopping-new}
\end{equation}
where in $t_2\!=\!0.21$\,eV, $t_3\!\approx\!0.05\,$eV are extracted from the Slater-Koster hopping parameters (see Eq.~(\ref{eq:hopping})). In order to be consistent with the choice of $t_2$ and $t_3$, we set $t_{\perp}\!=\!0.48\,$eV, which is also from the Slater-Koster formula (Eq.~(\ref{eq:hopping}). The phase factor $f(\k)\!=\!(e^{-i\sqrt{3}ak_y/3}+e^{i(k_xa/2+\sqrt{3}ak_y/6})+e^{i(-k_xa/2+\sqrt{3}ak_y/6})$. The interlayer hopping with $-$ stacking chirality $h_{-}\!=\!h_{+}^{\dagger}$.
The bandstructures of (2+2)-layer TMG at $\theta\!=\!1.05^{\circ}$ with the new interlayer hopping term Eq.~(\ref{eq:chiral-hopping-new}) are shown in Fig.~\ref{fig2b}, where (a) and (b) denote the cases with the same and opposite stacking chiralities respectively. Clearly the two low-energy bands marked by the red lines become more dispersive due to the presence of the further-neighbor interlayer hoppings, but the bandwidths are still small $\sim\!10-15$\,meV, and the two low-energy bands are still separated from the high-energy bands.

\section{The Chern-number hierarchy and quantum valley Hall effect}
\label{sec:qvh-chern}

\subsection{The Chern-number hierarchy}
\label{subsec:chern}
\begin{figure}
\includegraphics[width=3.5in]{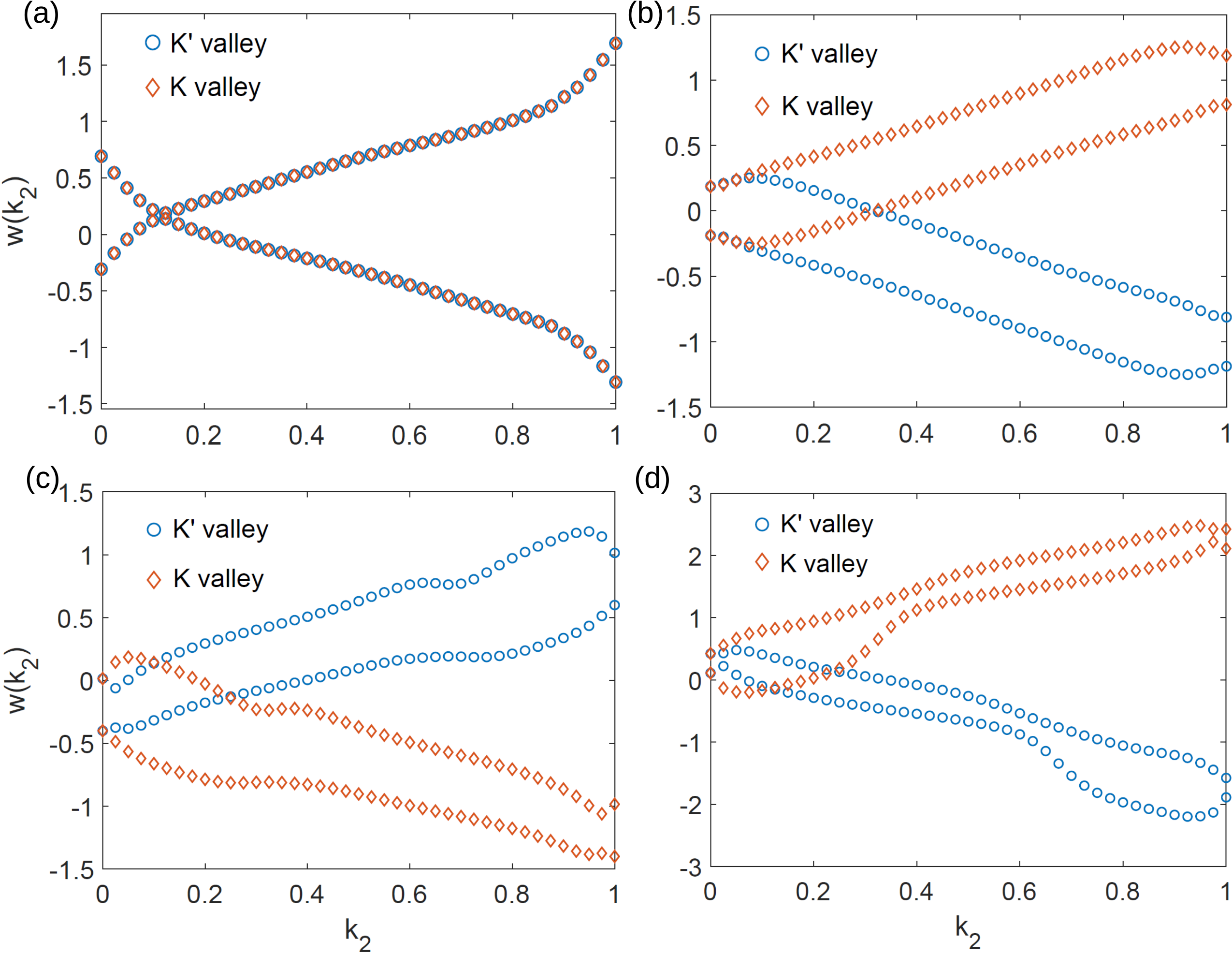}
\caption{The Wilson loops of $(M+N)$-layer twisted multilayer graphene at the first magic angle $\theta\!=\!1.05^{\circ}$: (a)$M\!=\!2, N\!=\!2$ with the same stacking chiralities; (b)$M\!=\!2, N\!=\!2$, with opposite stacking chiralities; (c) $M\!=\!2, N\!=\!4$, with the same stacking chiralities; (d) $M\!=\!2, N\!=\!4$, with opposite stacking chiralities.}
\label{fig3}
\end{figure}

The flat bands at the universal magic angle make the TMG systems a perfect platform to study the strongly correlated physics. In addition to the flat bands and the universal magic angles, the low-energy bands in the TMG systems also exhibit unusual topological properties with non-vanishing valley Chern numbers. To be specific, when the stacking chiralities of the $M$ layers and the $N$ layers are the same ,  the total Chern number of the two low-energy bands for each monolayer valley equals to $\pm(M-N)$. On the other hand, if the stacking chiralities of the $M$ layers and the $N$ layers are opposite, then the total Chern number of the two flat bands for each valley equals to  $\pm(M+N-2)$.  Such a Chern-number hierarchy is more concisely summerized in the following equation
\begin{align}
C_{\alpha,\alpha'}^{K}=+(\alpha(M-1)-\alpha'(N-1))\;,\nn
C_{\alpha,\alpha'}^{K'}=-(\alpha(M-1)-\alpha'(N-1))\;,
\label{eq:chern-number}
\end{align}
where  $C_{\alpha,\alpha'}^{K}$ ($C_{\alpha,\alpha'}^{K'}$) denotes the  total Chern number of the two low-energy flat bands for the $K$ ($K'$) valley, and the subscripts $\alpha, \alpha'=\pm$ represent the stacking chiralities of the $M$ layers and $N$ layers. We would like to emphasize that the total Chern number of the two flat bands (per valley per spin) is a more robust quantity than the Chern number of each individual flat band. This is because the former is protected by the energy gaps between the two flat bands and the other high-energy bands, while the latter is crucially dependent on how the gap between the two flat bands is opened up.

\begin{figure}
\includegraphics[width=3.5in]{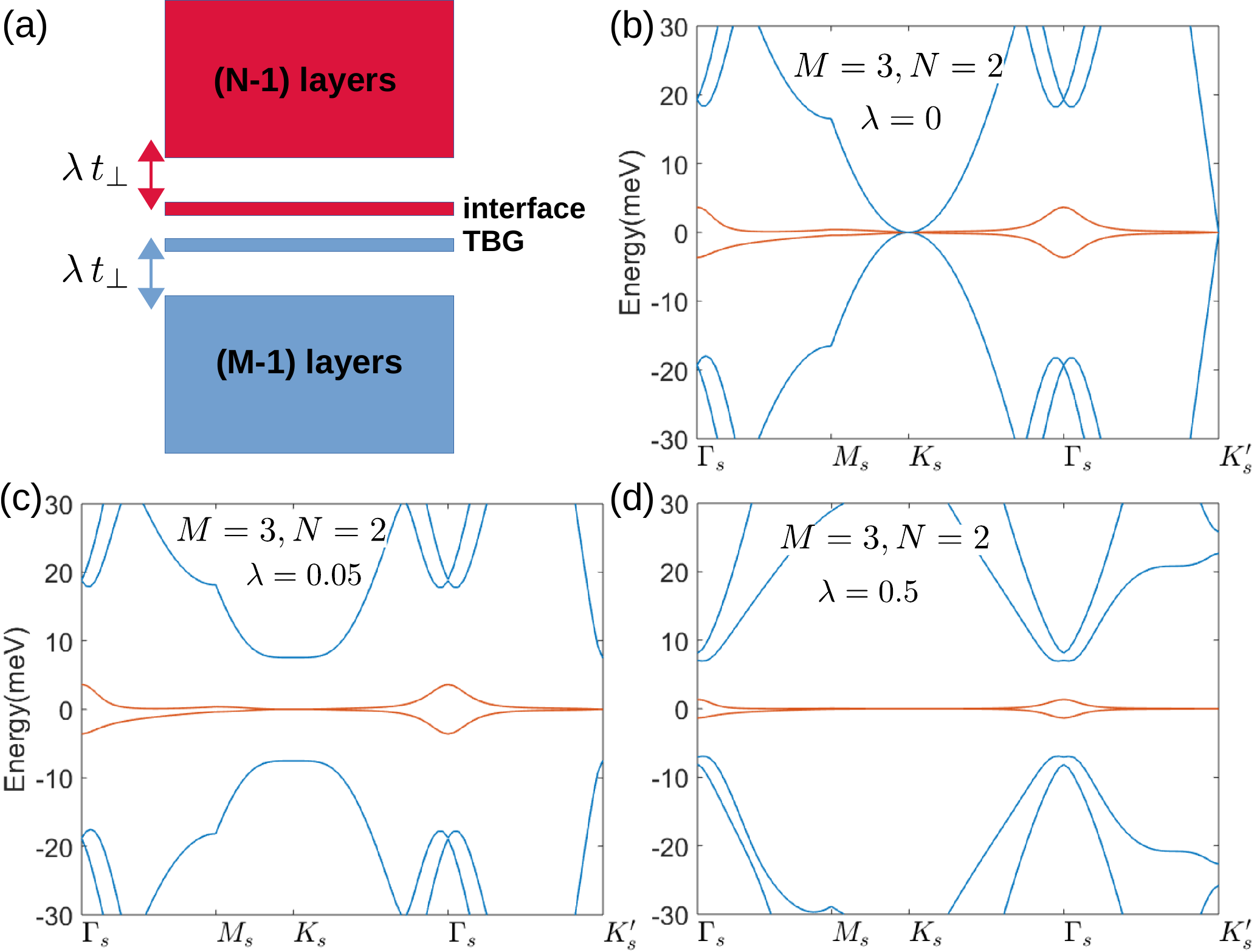}
\caption{(a) A schematic illustration of the TMG system. (b)-(d)  The bandstructures of $(2+3)$-layer TMG at $\theta\!=\!1.05^{\circ}$. The coupling parameter between the interface and the top (bottom) multilayers has been rescaled by $\lambda$: (b) $\lambda\!=\!0$, (c) $\lambda\!=\!0.05$,  and (d) $\lambda\!=\!0.5$.}
\label{fig4}
\end{figure}

In order to understand the Chern-number hierarchy of Eq.~(\ref{eq:chern-number}), we first divide the $(M+N)$-layer TMG system into three mutually decoupled subsystems: the TBG at the interface, the $(M-1)$  graphene monolayers below the interface TBG, and the $(N-1)$ graphene monolayers above the interface TBG, which are schematically shown in Fig.~\ref{fig4}(a).  We introduce a scaling parameter $0\!\leq\!\lambda\!\leq\!1$, and let the coupling strength between the three subsystems $t_{\perp}\!\to\!\lambda t_{\perp}$.  We adiabatically turn on the coupling between the three subsystems by increasing $\lambda$ from 0 to 1, then inspect the evolution of the bandstructures of the TMG system. 

In Fig.~\ref{fig4}(b) we show the bandstructure of $(3+2)$-layer TMG (of the $K$ valley) at $\theta\!=\!1.05^{\circ}$  with the scaling parameter $\lambda\!=\!0$. When $\lambda\!=\!0$, the magic-angle TBG at the interface would give rise to two flat bands with total Chern number 0 as marked by the red lines in Fig.~\ref{fig4}(a). The $(M-1)$ graphene monolayers below the TBG interface would contribute two low-energy bands with dispersions $\sim\!\pm\vert\k\vert^{M-1}$ around $K_s$
\footnote{The Dirac points of the $M$ graphene layers $K_M$ and of the $N$ graphene monolayers $K_N$  are respectively mapped to $K_s$ and $K_s'$ points of the moir\'e Brillouin zone, see Fig.~\ref{fig1}(b)}
.
Similarly, the $(N-1)$ layers above the TBG interface would contribute two low-energy bands with dispersions $\sim\!\pm\vert\k\vert^{N-1}$ around $K_s'$. Since we have considered the case $M\!=\!3$ and $N\!=\!2$,  there are quadratic band touching at $K_s$ and linear band touching at $K_s'$ in Fig.~\ref{fig4}(b). 
If $\lambda$ becomes nonzero, gaps would be opened up at $K_s$ and $K_s'$ for the bands with $\pm\vert\k\vert^{M-1}$ and $\pm\vert\k\vert^{N-1}$ dispersions as shown in Fig.~\ref{fig4}(c) for $\lambda\!=\!0.05$. As a result, the loop integral for the Berry's connection of the conduction and valance bands around a loop enclosing the $K_s$ point would acquire the \textit{same} Berry phase of $-\alpha (M-1)\pi$ ($\alpha$ is the stacking chirality of the $(M-1)$ layers), and
contribute $-\alpha(M-1)/2$ to the total Chern number respectively, which adds up to $-\alpha(M-1)$ (see Appendix \ref{sec:append-chern}). On the other hand, the conduction and valence bands around $K_s'$ point contributed by the $(N-1)$ layers above the interface would acquire the \textit{same} Berry phase of $\alpha'(N-1)\pi$ ($\alpha'$ is the stacking chirality of the $(N-1)$ layers), with the total Chern number of $\alpha'(N-1)$ (see Appendix \ref{sec:append-chern}). It is well known that the total Chern number of the bands from the $(M-1)$ and $(N-1)$ layers must cancel that of the two flat bands from the interface TBG, it follows that the total Chern number of the two flat bands for the $K$ valley equals to $\alpha(M-1)-\alpha' (N-1)$.
As $\lambda$ is further increased, the conduction and valence bands from the $(M-1)$ and $(N-1)$ layers are further pushed to high energies as shown in Fig.~\ref{fig4}(d) for $\lambda\!=\!0.5$, and the Chern number of the two flat bands would remain unchanged. Thus Eq.~(\ref{eq:chern-number}) has been proved. We refer the readers to Appendix \ref{sec:append-chern} for more details.

Eq.~(\ref{eq:chern-number}) has been numerically verified using the effective Hamiltonian of TMG shown in Eq.~(\ref{eq:HMN}). In particular, in Fig.~\ref{fig3}(a) we plot the Wilson-loop eigenvalues (denoted as $w(k)$) of the $(2+2)$ TMG ($M\!=\!2$, $N\!=\!2$) at the first magic angle  with the same stacking chirality ($\alpha\!=\!\alpha'\!=\!+$). The red diamonds and blue circles represent the Wilson loops of the $K$ and $K'$ valleys respectively. As clearly shown in the figure, for each valley the total Chern number of the two flat bands vanishes. In Fig.~\ref{fig3}(b) we  plot the Wilson loops of the $(2+2)$ TMG at the first magic angle, but with opposite stacking chiralities ($\alpha\!=\!+, \alpha'\!=\!-$). It is clearly seen that for the $K$ valley (blue circles) the two Wilson loops carry the same Chern number $+1$, giving rise to a total Chern number of $+2$ for the $K$ valley ($-2$ for the $K'$ valley), which is consistent with Eq.~(\ref{eq:chern-number}).  In Fig.~\ref{fig3}(c) and (d) we plot the Wilson-loop eigenvalues for the the $(2+4)$ TMG ($M\!=\!2$, $N\!=\!4$) at the first magic angle. When the stacking chiralities are the same, the total Chern number of the two flat bands for the $K$ ($K'$) valley equals to $-2$ ($+2$); while if the stacking chiralities are opposite, then the total Chern number of the two flat bands equals to $+4$ ($-4$) for the $K$ ($K'$) valley. Again, this is in perfect agreement with Eq.~(\ref{eq:chern-number}). We have also numerically tested the other $(M+N)$-layer TMG with $M, N$ extending from 1 to 5, and they are all consistent with Eq.~(\ref{eq:chern-number}).

We have also considered the more realistic situation in which the interlayer hopping is given by Eq.~(\ref{eq:chiral-hopping-new}) instead of Eq.~(\ref{eq:chiral-hopping}). Since the chiral symmetry is broken in Eq.~(\ref{eq:chiral-hopping-new}), the Chern-number hierachy given by Eq.~(\ref{eq:chern-number}) is no longer exact. However, since the Chern number of concern is the total Chern number of the two low-energy flat bands, it should remain unchanged as long as the two low-energy flat bands remain separated from the other high-energy bands. We have numerically calculated the total valley Chern numbers of the two low-energy bands in $(M+N)-$layer TMG systems ($M, N$ varies from 1 to 5) at $\theta\!=\!1.05^{\circ}$ using the more realistic interlayer hopping term Eq.~(\ref{eq:chiral-hopping-new}), and find that the Chern-number hierarchy of Eq.~(\ref{eq:chern-number}) remains  correct for the cases of $(M=1,N=2)$, $(M=1,N=3)$, $(M=1,N=4)$, $(M=1,N=5)$, $(M=2,N=2)$, $(M=2,N=3)$, and is partially correct (for one stacking configuration) for $(M\!=\!3,N\!=\!3)$, $(M\!=\!2, N\!=\!4)$, and $(M\!=\!3, N\!=\!5)$.

\subsection{Gate tunable quantum valley Hall effect }
The valley Chern numbers given by Eq.~(\ref{eq:chern-number}) can be further tuned by applying an vertical electric potential $V_{\perp}$. Taking the case of $(2+2)$-layer TMG and $(2+1)-$layer TMG as an example, we study the dependence of the valley Chern numbers of each of the two low-energy bands on the vertical electric potential $V_{\perp}$ at $\theta\!=\!1.05^{\circ}$. The valley Chern number of each band is denoted by $C_{n\alpha\alpha'}^{K}$, with the band index $n=1,2$, the stacking chirality $\alpha,\alpha'\!=\!\pm$, and the valley index $K$.

In Table ~\ref{table:chern-number-1} we show the dependence of $C_{1\alpha\alpha'}^{K}$, $C_{2\alpha\alpha'}^{K}$ on the vertical electric potential $V_\perp$ (in units of meV) for  $(2+2)$-layer TMG. The valley Chern numbers are calculated using the effective Hamiltonian Eq.~(\ref{eq:HMN}) with the more realistic interlayer hopping Eq.~(\ref{eq:chiral-hopping-new}). When the stacking chiralities are the same (both $+$), the Chern number of each of the two low-energy bands becomes $\pm 3$ once a small $V_{\perp}\!\sim\!12\,$meV is applied, whereas the total Chern number of the two bands still sums to zero. As $\vert V_{\perp}\vert$ increases, the valley Chern number of the two flat bands are changed to $\pm 2$ at $\vert V_{\perp}\vert\!=\!24\,$meV, then become $0$ and $\pm 1$ respectively at $\vert V_{\perp}\vert\!=\!36\,$meV, and both become $\pm1$ when $\vert V_{\perp}\vert\gtrapprox 48\,$meV.  On the other hands, when the stacking chiralities are opposite ($\alpha=+$ and $\alpha'=-$), the total valley Chern number of the two bands equals to $-2$ at $\vert V_{\perp}\vert\!=\!0$, and remains unchanged for $\vert V_{\perp}\vert\lessapprox\!40\,$meV. Then the total valley Chern number of the two bands becomes $+1$ for $\vert V_{\perp}\vert\gtrapprox\!48\,$meV. Table~\ref{table:chern-number-1} indicates that the topological phases of $(2+2)-$layer TMG is highly tunable by gate voltage, which is qualitatively in agreement with the results reported in Refs.~\onlinecite{zhang-tbg-prb19} and \onlinecite{ashvin-double-bilayer-arxiv19}. Moreover, here we have also shown that it is sensitive to the stacking configurations.

In Table~\ref{table:chern-number-2}  we show  $C_{n+}^{K}$ ($n=1,2$) 
 \textit{vs.} $V_{\perp}$ for $(2+1)$-layer TMG at $\theta\!=\!1.05^{\circ}$, where the subscript ``$+$" means that the bottom bilayer has $+$ stacking chirality, and $n=1,2$ is the band index. 
 When the vertical potential $V_\perp\!=\!0$, $C_{1+}^{K}\!=\!+1$ and $C_{2+}^{K}\!=\!0$, which is consistent with Eq.~(\ref{eq:chern-number}).   Once a small positive or negative $V_{\perp}\!\sim\!10\,$meV is applied, $C_{1+}^{K}$ is changed to $+2$ and $C_{2+}^{K}$ becomes -1. However, when $\vert V_{\perp}\vert\!\gtrapprox\!20\,$meV, the valley Chern numbers become highly dependent on the sign of $V_{\perp}$ due to the asymmetric stacking configuration of $(2+1)$-layer TMG. Thus distinct topological phases could be realized by reversing the gate potentials.

\begin{table}[bth]
\caption{The dependence of the Chern numbers on vertical electric potential for (2+2)-layer TMG}
\begin{ruledtabular}
\begin{tabular}{lclclclclclc}
$V_{\perp}$& -60 & -48 & -36& -24 & -12 & 0 & 12 & 24 & 36 & 48 & 60 \\
$C^{K}_{1++}$ & +1 & +1 & 0 & -2 & -3  &  -  & +3 & +2 & 0 &-1 &-1 \\
$C^{K}_{2++}$ & +1 & +1 & +1 & +2  & +3 & -&-3 &-2  & -1 & -1 & -1 \\
$C^{K}_{1+-}$ & 0 &  0 & +1 & +1 &  0 & 0 & 0 & +1 & +1  &  0 & 0\\
$C^{K}_{2+-}$ & +1& +1 & +1 & +1 & +2 & +2 &+2 & +1& +1  & +1 & +1
\end{tabular}
\end{ruledtabular}
\label{table:chern-number-1}
\end{table}
\begin{table}[bth]
\caption{The dependence of the Chern numbers on vertical electric potential for (2+1)-layer TMG}
\begin{ruledtabular}
\begin{tabular}{lclclclclclc}
$V_{\perp}$& -60 & -48 & -36& -24 & -12 & 0 & 12 & 24 & 36 & 48 & 60 \\
$C^{K}_{1+}$ & 0 &  -1 & -1  & -1 & +2 & +1 & +2 & +2 & +2 &+2 & +2 \\
$C^{K}_{2+}$ & +2 & +2 & +2 & +2  & -1  &  0  & -1  & -1  & -1 & -1 & -1
\end{tabular}
\end{ruledtabular}
\label{table:chern-number-2}
\end{table}

\section{Verifications using an empirical tight-binding model}
\label{sec:tbmodel}

The flat bands at the first magic angle (Fig.~\ref{fig2}) and the Chern-number hierarchy (Eq.~(\ref{eq:chern-number})) in TMG have been verified using a realistic microscopic tight-binding (TB) model. To be specific, the hopping parameter between two $p_z$ orbitals at different carbon sites $i$ and $j$ in the multilayer system is expressed in the Slater-Koster form
\begin{equation}
t(\mathbf{d})=V_{\sigma}\,(\frac{\mathbf{d}\cdot\mathbf{\hat{z}}}{d})^2+V_{\pi}\,(\,1-(\frac{\mathbf{d}\cdot\mathbf{\hat{z}}}{d})^2\,)
\label{eq:hopping}
\end{equation}
where  $V_{\sigma}=V_{\sigma}^{0}\,e^{-(r-d_c)/\delta_0}$, and
$V_{\pi}=V_{\pi}^{0}\,e^{-(r-a_0)/\delta_0}$. $\mathbf{d}=(d_x, d_y, d_z)$ is  the displacement vector between the two carbon sites. $a_0=a/\sqrt{3}=1.42\,$\angstrom, $d_c=3.35\,$\angstrom\ is the interlayer distance in AB-stacked bilayer graphene, and $\delta_0=0.184\,a$. $V_{\sigma}^{0}=0.48\,$eV and $V_{\pi}^{0}=-2.7\,$eV. The atomic corrugations at the interface between the $M$ layers and the $N$ layers in $(M+N)$ TMG are modeled by Eq.~(\ref{eq:dz-1}), and their effects can be taken into account by plugging Eq.~(\ref{eq:dz-1}) into the hopping parameter shown in Eq.~(\ref{eq:hopping}). 

\begin{figure}
\includegraphics[width=3.5in]{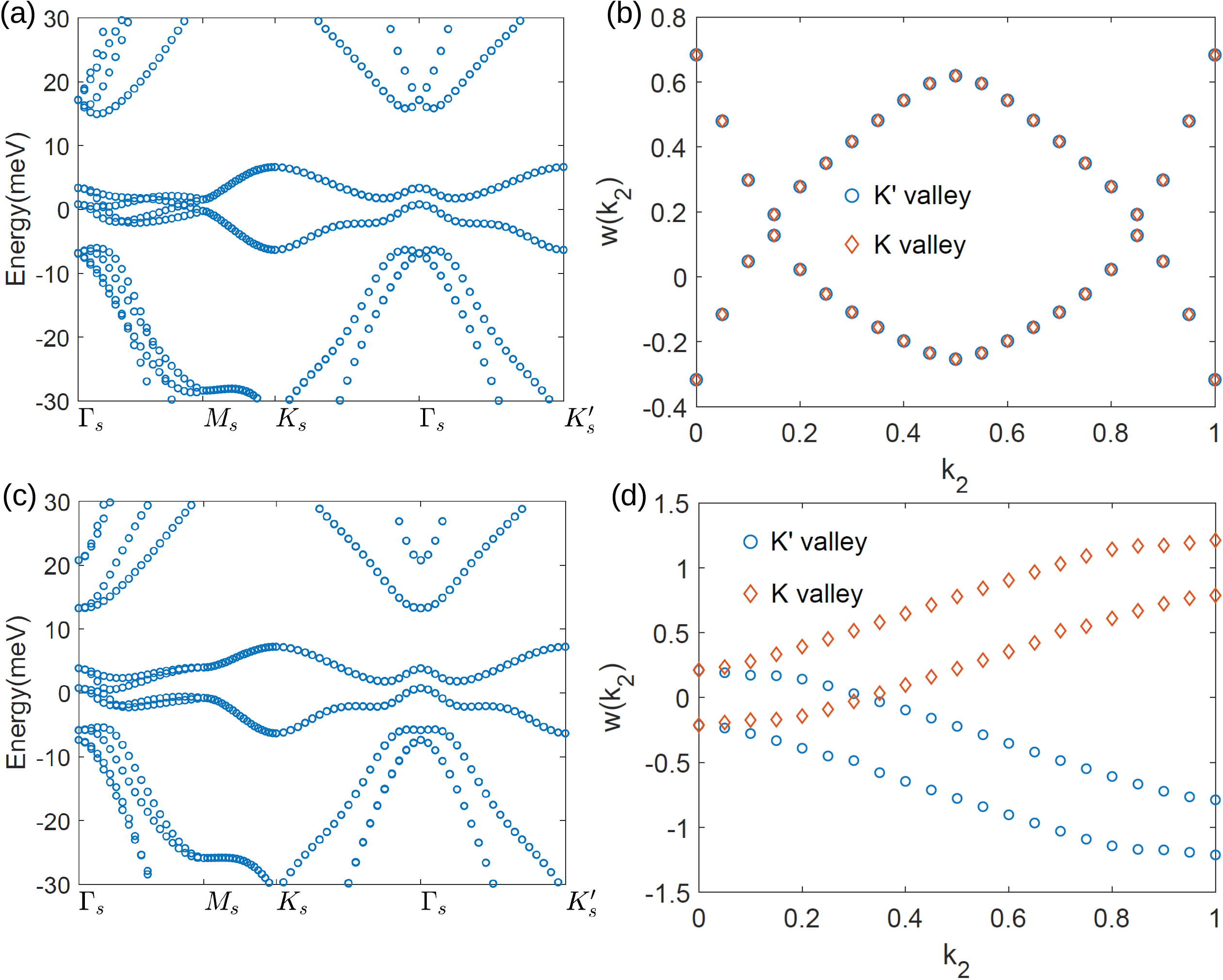}
\caption{(The bandstructures of (2+2) TMG at $\theta\!\approx\!1.08^{\circ}$: (a) with opposite stacking chiralities, and (c) with the same stacking chirality. The Wilson loops of (2+2) TMG at $\theta\!\approx\!1.08^{\circ}$: (b) with opposite stacking chiralities, and (d) with the same stacking chirality. The blue circles and red diamonds denote the Wilson loops of the $K$ and $K'$  valleys.}
\label{fig5}
\end{figure}

The bandstructures calculated using the Slater-Koster TB model at $\theta\!\approx\!1.08^{\circ}$ for the $(2+2)$ TMG are shown in Fig.~\ref{fig4}.  To be specific, the bandstructure of (2+2) TMG with opposite and the same stacking chiralities are shown in Fig.~\ref{fig4}(a) and (c) respectively. It is evident that there are four low-energy bands (contributed by the two valleys) which are separated from the other high-energy bands, and the bandwidths are on the order of 10-15\,meV, which  are greater than those from the continuum model (see Fig.~\ref{fig2}(a). This is because in the continuum model only the nearest neighbor interlayer hopping is kept  (see Eq.~(\ref{eq:H22t})), which imposes a chiral symmetry to the Hamiltonian of Eq.~(\ref{eq:HMN}). As argued in Appendix \ref{sec:append2}, the chiral symmetry of Eq.~(\ref{eq:HMN}) would pin the zeroth pseudo LLs emerging from the twisted bilayer at the interface to zero energy, leading to almost vanishing bandwidth as shown in Fig.~\ref{fig2}. However, such a chiral symmetry is broken in the realistic Slater-Koster TB model, and the bandwidth of the flat bands is expected to be enhanced due to the presence of further neighbor interlayer hoppings. 

In Fig.~\ref{fig5}(b) and (d) we show the Wilson loops of the four low-energy bands of (2+2) TMG at $\theta\!\approx\!1.08^{\circ}$ calculated using the Slater-Koster TB model. Fig.~\ref{fig5}(b) ((d)) denotes the case with opposite (the same) stacking chiralities with the valley Chern number being $\pm 2$ ($0$). 
The topological equivalence between the band structures obtained by the more accurate tight binding model as shown in Fig.~\ref{fig5} and by the continuum model as shown in Fig.~\ref{fig3}(a)-(b) provides a strong supporting fact for the Chern number hierarchy given by Eq.~(\ref{eq:chern-number}).

\section{Valley-contrasting orbital magnetizations and valley-polarized states}
\begin{figure}
\includegraphics[width=3.5in]{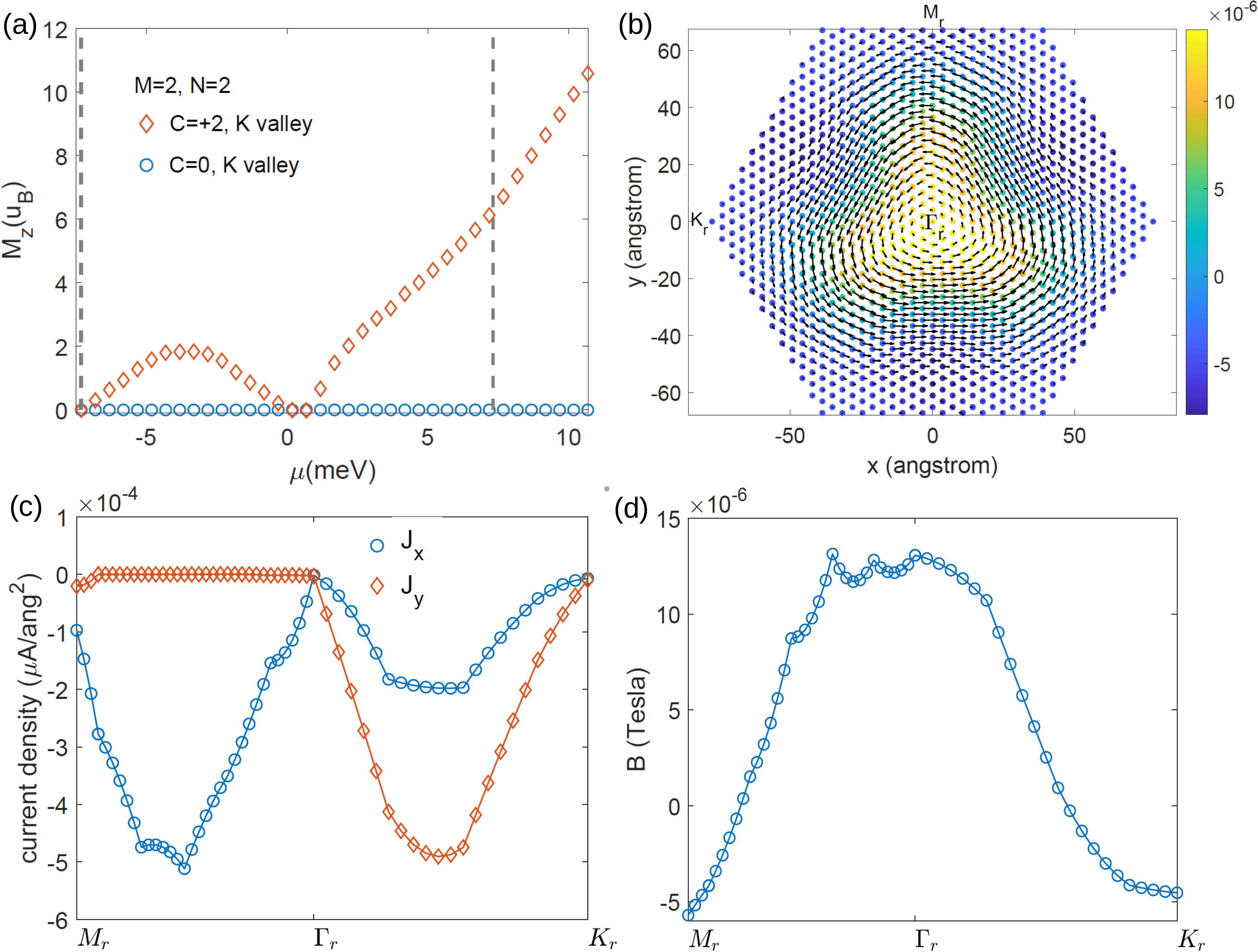}
\caption{(a) The orbital magnetization $M_z$ of the two flat bands of the $K$ valley for $(2+2)$-layer twisted graphene at the first magic angle. The red circles (blue circles) represent the situation that the stacking chiralities are opposite (the same), with the Chern number +2 (0).  The horizontal axis is the chemical potential $\mu$. The vertical dashed lines mark the bulk band edges of the two low-energy flat bands. (b) The distributions of the current density (black arrows) and the current-induced magnetic field (color coding) within the moir\'e supercell. The $AA$ region is centered at the origin.
(c) Current densities plotted along the real-space path $M_r-\Gamma_r-K_r$, where the $M_r$, $\Gamma_r$, and $K_r$ points are marked in (b). (d) The local magnetic field plotted along  $M_r-\Gamma_r-K_r$. }
\label{fig6}
\end{figure}

The valley Chern numbers given by Eq.~(\ref{eq:chern-number}) implies opposite orbital magnetizations for the two monolayer valleys $K$ and $K'$. In particular, according to the ``modern theory" of orbital magnetizations \cite{orbital-mag-prl05,xiao-dos-prl05,orbital-mag-prb06,orbital-mag-prl07}, the orbital magnetization of the $(M+N)$-layer TMG for  either $K$ or $K'$ valley can be expressed as 
\begin{align}
M_z
=\frac{e}{2\hbar(2\pi)^2}&\sum_n\int_{\epsilon_{n\k}\leq\mu} d\mathbf{k}\,\textrm{Im}\{\langle\partial_{\k}u_{n\k}\vert\;\nn
&\times\,(H_{\k}+\epsilon_{n\k}-2\mu)\,\vert \partial_{\k}u_{n\k}\rangle\}\;,
\label{eq:mz}
\end{align}
where $\epsilon_{n\k}$ and $\vert u_{n\k}\rangle$ are the eigenenergies and  (the periodic part of) Bloch eigenstates of a $(M+N)$-layer TMG Hamiltonian (denoted by $H$) for either $K$ or $K'$ valley,  $\mu$ is the chemical potential, and $H_{\k}\!=\!e^{-i\k\cdot\mathbf{r}} H e^{i\k\cdot\mathbf{r}}$. Since the $K$ and $K'$ valleys are transformed to each other by a time-reversal operation, it is naturally expected that the valley-contrasting Chern numbers shown in Eq.~(\ref{eq:chern-number}) would lead to opposite orbital magnetization for the two valleys.  The valley-contrasting orbital magnetizations further suggests that  when the valley degeneracy is lifted by external magnetic fields or by Coulomb interactions, a valley-polarized state with non-vanishing or even quantized anomalous Hall conductivity will be generated due to the nonzero valley Chern numbers.

We have exploited this idea using the continuum model given by Eq.~(\ref{eq:HMN}) with the interlayer hopping modeled by Eq.~(\ref{eq:chiral-hopping-new}). To be explicit, we consider the case of $(2+2)$-layer TMG at the first magic angle $\theta\!=\!1.05^{\circ}$. We have calculated the orbital magnetizations ($M_z$) of the two low-energy bands for the $K$ valley using  the  Hamiltonian Eq.~(\ref{eq:HMN}) with the interlayer hopping modeled by Eq.~(\ref{eq:chiral-hopping-new}). The dependence of $M_z$ on the chemical potential $\mu$ is plotted in Fig.~\ref{fig6}(a).  The red  diamonds in Fig.~\ref{fig6}(a) represent the situation that the bottom bilayer and  the top bilayer have opposite stacking chiralities with the valley Chern number $\pm2$ (see Eq.~(\ref{eq:chern-number})). In this case the magnitude of $M_z$ is large, which is on the order of $10\,\mu_B$ per moir\'e supercell when the two flat bands are completely filled. The large orbital magnetization is a manifestation of the band topology on the moir\'e pattern. In particular,
the non-vanishing Chern number of the $K$ or $K'$ valley implies that the ground state at a given filling would possess chiral current loops. The characteristic radius of the current loop is on the order of the moir\'e length scale $L_s$, which is associated with large orbital angular momentum $L_z\sim \mathbf{r}\times \mathbf{p}$, with $\vert\mathbf{p}\vert\!\sim\!\hbar K$ and $\vert\mathbf{r}\vert\!\sim\!L_s\!\gg\!a$. Therefore, the orbital magnetization generated by the current loops on the moir\'e scale is expected to be much greater than that on the microscopic lattice scale. We also note that $M_z$ increases almost linearly with $\mu$ when $\mu$ is in the gap above the two flat bands, which is a signature of the non-vanishing Chern number \cite{orbital-mag-prb06}.  
On the other hand, the blue circles in Fig.~\ref{fig6}(a) denote the case that the bottom bilayer and the top bilayer have the same stacking chirality with vanishing valley Chern number. In this case the orbital magnetization vanishes identically for any chemical potential due to the presence of a $\mathcal{P}\mathcal{T}$ symmetry ($\mathcal{P}$ is the 3D spatial inversion operation). 
It worth to note again that the orbital magnetization plotted in Fig.~\ref{fig6}(a) is for the $K$ valley. The orbital magnetization for the $K'$ valley is just opposite to that of the $K$ valley.

The large orbital magnetizations imply that the valley degeneracy of the system can be easily lifted by a weak external magnetic field or by spontaneous valley symmetry breaking from Coulomb interactions, leading to a valley-polarized (quantum) anomalous Hall state.  
A rough estimate reveals that a magnetic field of $2\,$T would give rise to an orbital (or valley) Zeeman splitting $\sim 2\,$meV (15\% of the bandwidth), which would lead to a state with considerable valley polarization and anomalous Hall effect. Such a valley polarized anomalous Hall state is expected to possess chiral current loops, which are responsible for the large orbital magnetization shown in Fig.~\ref{fig6}(a).  Here we assume that the $K$ valley is 100\% polarized either due to the presence of an external magnetic field, or due to spontaneous valley symmetry breaking  from Coulomb interactions, and we calculate the local current density  $\mathbf{J}(\mathbf{r})$ and the current-induced local magnetic field $B(\mathbf{r})$ with the two flat bands of the $K$ valley being completely filled. In Fig.~\ref{fig6}(b) we show the distributions of $\mathbf{J}(\mathbf{r})$ and $B(\mathbf{r})$ within the moir\'e Wigner-Seitz cell for the $(2+2)$-layer TMG with opposite stacking chiralities, and with 100\% valley polarization.  The small filled circles in Fig.~\ref{fig6}(b) represent the discretized real-space positions, with the color coding denoting the strength of the local magnetic field in units of Tesla. The black arrows represent the local current densities whose amplitudes are proportional to the lengths of the arrows. Clearly the valley polarized ground state possesses chiral current loops circulating around the $AA$ region.  These circulating current loops further generate magnetic fields in the $AA$ region with the magnitude $\sim\!10^{-5}\,$T, which may be a strong experimental evidence for the non-vanishing valley Chern number in $(2+2)$-layer TMG with opposite stacking chiralties.  

In Fig.~\ref{fig6}(c) we plot the current densities $J_x$ (blue circles) and $J_y$ (red diamonds) along the real-space path $M_r-\Gamma_r-K_r$ in units of $\mu \textrm{A}/\angstrom^2$, where the $M_r$, $\Gamma_r$, and $K_r$ points are marked in Fig.~\ref{fig6}(b). It is interesting to note that $J_y$ almost vanishes identically along the $M_r-\Gamma_r$ path due to the winding pattern of the current. The maximal magnitude of the current density $\sim 6\times 10^{-4}\mu \textrm{A}/\angstrom^2$. In Fig.~\ref{fig6}(d) we show the local magnetic field plotted along the $M_r-\Gamma_r-K_r$ path. Clearly the magnetic field has a peak centered at $\Gamma_r$ (the $AA$ point) with the magnitude $\sim\!10^{-5}$\,T.
The details of the computing the current densities and local magnetic fields are presented in Appendix \ref{append:current}.

\section{Summary}
To summarize,we have studied the electronic structures and topological properties of the $(M+N)$-layer TMG system. We have proposed that, with the chiral approximation, there always exists two low-energy bands whose bandwidths become minimally small at the magic angle of TBG. We have further shown that the two flat bands in the TMG system are topologically nontrivial, and exhibits a Chern-number hierarchy. In particular, when the stacking chiralities of the $M$ layers and the $N$ layers are opposite,  the total Chern number of the two low-energy bands for each valley equals to $\pm(M+N-2)$ (per spin). On the other hand, if the stacking chiralities of the $M$ layers and the $N$ layers are the same, then the total Chern number of the two low-energy bands for each valley is  $\pm(M-N)$ (per spin). The non-vanishing valley Chern numbers are associated with large and valley-contrasting orbital magnetizations along $z$ direction, which implies that the valley degeneracy can be lifted by weak external magnetic fields or by Coulomb interactions, leading to a valley-polarized anomalous Hall state. Such a valley polarized state is associated with chiral current loops circulating around the $AA$ region, which generates local magnetic fields peaked at the $AA$ region. The local magnetic fields induced by the chiral current loops may be a robust experimental signature of the valley polarized state with non-vanishing Chern number.
Our work is a crucial step forward in understanding the electronic properties of twisted multilayer graphene. The universal magic angles and the Chern-number hierarchy proposed in this work make the TMG system a perfect platform to study the interplay between  electrons' Coulomb correlations and nontrivial band topology.

\textit{Note added}: During the preparation of our manuscript, we note two recently posted works Ref.~\onlinecite{ashvin-double-bilayer-arxiv19} and  Ref.~\onlinecite{koshino-double-bilayer-arxiv19} . In the former, the electronic structures, superconductivity, and correlated insulating phase have been discussed for twisted double bilayer graphene with the same stacking chirality ($AB$-$AB$ stacking). In the latter, the bandstructures and valley Chern numbers of twisted double bilayer graphene with different stacking chiralities have been discussed.

\acknowledgements
J.L. and  X.D. acknowledge  financial support from the Hong Kong Research Grants Council (Project No. GRF16300918). We thank Hongming Weng for invaluable discussions. 

\appendix

\section{The flat bands and  magic angles in the TMG system}
\label{sec:append2}
In this appendix we explain the origin of the flat bands and universal magic angles in the $(M+N)$-layer TMG system. After some gauge transformations, the constant wavevectors $\mathbf{K}_{M}, \mathbf{K}_N$ in Eq.~(\ref{eq:HM}) can be removed. Taking the case of $(2+2)$-layer TMG with the same stacking chiralities as an example,  the effective Hamiltonian is explicitly written as (after the gauge transformation)
\begin{equation}
H_{+,+}^{K}(2+2)=\begin{pmatrix}
h_0(\hat{\k}) & h_{+} & 0 & 0 \\
h_{+}^{\dagger} & h_0(\hat{\k}) & U(\mathbf{r}) & 0 \\
0 & U^{\dagger}(\mathbf{r}) & h_0(\hat{\k}) & h_{+} \\
0 & 0 & h_{+}^{\dagger} & h_0(\hat{\k})
\end{pmatrix}\;,
\label{eq:H22}
\end{equation}
where $h_0(\hat{\k})=-\hbar v_F\hat{\k}\!\cdot\!\mathbf{\sigma}$, and $h_{\alpha}$ and $U$ are defined in Eq.~(\ref{eq:chiral-hopping}) and Eq.~(\ref{eq:twist-coupling1}) of the main text. Then we make the following unitary transformation to the basis functions of Eq.~(\ref{eq:H22}) (and more generally, to those of  Eq.~(\ref{eq:HMN}) of the main text)
\begin{align}
&\vert\widetilde{\psi}_{M s}\rangle=\frac{1}{2}\Big(\,\vert{\psi}_{Ms}\rangle+i\vert{\psi}_{(M+1) s}\rangle\,\Big)\;,\nn
&\vert\widetilde{\psi}_{(M+1) s}\rangle=\frac{1}{2}\Big(\,\vert{\psi}_{Ms}\rangle-i\vert{\psi}_{(M+1) s}\rangle\,\Big)\;,\nn
&\vert\widetilde{\psi}_{l s}\rangle=\vert{\psi}_{l s}\rangle,   \textrm{  if $l\neq M, M+1$}
\label{eq:unitary-transform}
\end{align}, 
where $\vert{\psi}_{l s}\rangle$ denotes the Bloch state at the $K$ point from the $l$th layer and the $s$ sublattice. Applying the unitary transformation Eq.~(\ref{eq:unitary-transform}) to Eq.~(\ref{eq:H22}) (letting $M\!=\!2$), then expanding the interlayer coupling term $U(\mathbf{r})$ to the linear order of $r/L_s$,  Eq.~(\ref{eq:H22}) becomes
\begin{equation}
\widetilde{H}_{+,+}^{K}(2+2)=
\begin{pmatrix}
h_0(\hat{\k}) & \frac{h_{+}}{\sqrt{2}} & \frac{h_{+}}{\sqrt{2}} & 0 \\
\frac{h_{+}^{\dagger}}{\sqrt{2}} & h_0(\hat{\k}-\frac{e}{\hbar}\mathbf{A}) & -3iu_0 & -i\frac{h_{+}}{\sqrt{2}} \\
\frac{h_{+}^{\dagger}}{\sqrt{2}} & 3i u_0 & h_0(\hat{\k}+\frac{e}{\hbar}\mathbf{A}) & i\frac{h_{+}}{\sqrt{2}}  \\
0 & -i\frac{h_{+}^{\dagger}}{\sqrt{2}}  & i\frac{h_{+}^{\dagger}}{\sqrt{2}}  & h_0(\hat{\k})
\end{pmatrix}
\label{eq:H22t}
\end{equation}
where the pseudo vector potential $\mathbf{A}\!=\!(2\pi u_0')/(L_s ev_{F})\,(\,y\, , -x\,)$ \cite{liu-ll-arxiv}.  Note that the diagonal blocks $ h_0(\hat{\k}\pm\frac{e}{\hbar}\mathbf{A})\!=\!-\hbar v_F(\hat{\k}\pm e\mathbf{A}/\hbar)$ are equivalent to the Dirac fermions coupled with opposite pseudo magnetic fields,  which would generate pseudo LLs of opposite Chern numbers $\pm 1$ \cite{liu-ll-arxiv}. In particular, the two zeroth pseudo LLs have opposite sublattice polarizations, thus the intrasublattice coupling term  $\pm 3i u_0$ in Eq.~(\ref{eq:H22t}) vanishes in the subspace of the zeroth pseudo LLs \cite{liu-ll-arxiv}. Therefore, if we re-write Eq.~(\ref{eq:H22t}) in a \textit{mixed basis} consisted of the Dirac states from the first and fourth layers, and the zeroth pseudo LLs from the second and third layers, then $\widetilde{H}_{+,+}^{K}(2+2)$ becomes 
\begin{equation}
\widetilde{H}_{+,+}^{K}(2+2)\!\approx\!
\begin{pmatrix}
h_{b}  & 0 \\
0 & h_t 
\end{pmatrix}\;.
\label{eq:H22ll}
\end{equation}
Each element of Eq.~(\ref{eq:H22ll}) is a $3\times 3$ matrix.
In particular,
\begin{equation}
h_{b}=\begin{pmatrix}
0 & \hbar v_F(\hat{k}_x+i\hat{k}_y) & 0 \\
 \hbar v_F(\hat{k}_x- i\hat{k}_y) & 0  & t_{\perp}\eta(k) \\
 0 & t_{\perp}\eta^{*}(k) & 0
\end{pmatrix}\;,
\label{eq:hb}
\end{equation}
and 
\begin{equation}
h_{t}=\begin{pmatrix}
0 & it_\perp\eta(k) & 0 \\
-it_\perp\eta^*(k) & 0 &  \hbar v_F(\hat{k}_x+ i\hat{k}_y) \\
0 &  \hbar v_F(\hat{k}_x- i\hat{k}_y) & 0 
\end{pmatrix}\;,
\label{eq:ht}
\end{equation}
where $\eta(k)$ denotes the coupling between the zeroth pseudo LL and the Dirac states of the first (fourth) layer, which can be expressed as some integral over the eigenfunctions of the zeroth pseudo LLs and the Bloch functions, and $k$ is index indicating the zeroth Landau-level degeneracy.  Note that we have dropped the higher pseudo LLs in Eq.~(\ref{eq:H22ll}). 

The Diagonalizations of Eq.~(\ref{eq:ht}) and Eq.~(\ref{eq:hb}) would always yield two decoupled zero modes at any $k$. These two zero modes originate from the two zeroth pseudo LLs contributed by the $M$th and $(M+1)$th twisted bilayer ($M\!=\!2$ for Eq.~(\ref{eq:H22ll})), and they stay at zero energy even after being coupled with the other layers due to the chiral symmetry of the effective Hamiltonian of TMG: all the matrix elements in Eq.~(\ref{eq:HMN}) and Eq.~(\ref{eq:H22})-Eq.~(\ref{eq:H22ll}) are \textit{intersublattice}. As a consequence, if we apply the gauge transformation such that all the basis functions at the $B$ sublattice changes sign, then the total Hamiltonian and eigenenergies would change sign as well. However, the eigenenergies are supposed to be invariant under such a gauge transformation, which thus enforces that both $E(\k)$ and $-E(\k)$ have to be the eigenenergies of the Hamiltonian. Therefore, a zero mode would stay at zero energy as long as the chiral symmetry is preserved. Similar argument applies to any $(M+N)$-layer TMG systems with either opposite or the same stacking chiralities. As long as the chiral symmetry is preserved, the zeroth pseudo LLs emerging from the interface between the $M$ layers and $N$ layers would be pinned to zero energy. 

On the other hands, it is well known that at the magic angles of TBG the bandwidth of the two low-energy bands for each valley is minimal. From the perspective of the pseudo LLs \cite{liu-ll-arxiv}, it means that at the magic angles, the states within the zeroth pseudo LLs are minimally coupled with each other (and to the higher pseudo LLs), thus they are almost exactly flat. As discussed above, by virtue of the chiral symmetry, these zeroth pseudo LLs that are maximally flat at the magic angles would remain flat  even after being coupled with the other layers in the TMG systems. It  follows that the magic angles in TBG should be \textit{universal} in the TMG systems by virtue of the chiral symmetry in Eq.~(\ref{eq:HMN}).

\section{Derivations of the Chern-number hierarchy}
\label{sec:append-chern}
In this Appendix we mathematically prove the Chern-number hierarchy given by Eq.~(\ref{eq:chern-number}). As discussed in Sec.~\ref{subsec:chern}, the $(M+N)$-layer TMG system can be decomposed into three subsystems: the TBG at the interface, the $(M-1)$ layers below the interface, and the $(N-1)$ layers above the interface. This is schematically shown in Fig.~\ref{fig4}(a). In graphene multilayers with $+$ stacking chirality, the $B$  sublattice of the $n$th layer is strongly coupled with the $A$  sublattice of the $(n+1)th$ layer, forming pairs of bounding and anti-bounding states, leaving the $A$ sublattice of the first layer and the $B$ sublattice of the $N$th layer ($N$ is the number of layers) as two low-energy degrees freedom. Similarly, in graphene multlayers with $-$ stacking chirality,  the $B$ sublattice of the 1st layer and the $A$ sublattice of the $N$th layer would be the low-energy degrees of freedom. The low-energy effective Hamiltonians of the $(M-1)$ layer graphene around the $K$ valley with $\alpha$ ($\alpha=\pm1$) stacking chirality can be obtained by straightforward $(M-1)$th order perturbation theory \cite{min-tmg-08}, which  is expressed as 
\begin{align}
&H_{\textrm{eff}}^{\alpha}(M-1)\;\nn
=&(-1)^{M-2}\begin{pmatrix}
0  & \frac{(\hbar v_{F}(k_x+i\alpha k_y))^{M-1}}{t_{\perp}^{M-2}} \;\\
\frac{(\hbar v_{F}(k_x-i\alpha k_y))^{M-1}}{t_{\perp}^{M-2}} & 0 \;
\end{pmatrix}\;,
\label{eq:heff}
\end{align}
where $t_{\perp}$ is the interlayer hopping within the $(M-1)$ layers. Eq.~(\ref{eq:heff}) is in the basis of $\vert 1, A\rangle$ ($\vert 1, B\rangle$) and $\vert M-1,B\rangle$ ($\vert M-1, A\rangle$) if the $(M-1)$ layers have $+$ ($-$) stacking chirality.
On the other hand, there are two nearly flat bands at the magic angle contributed by the interface TBG. Around the $K_s$ and $K_s'$ points these two flat bands are equivalent to zeroth pseudo LLs, and possess opposite sublattice polarizations as argued in Ref.~\onlinecite{liu-ll-arxiv}. Let us first assume the coupling between the $(M-1)$ layers and the interface TBG is small,  then  the low-energy effective Hamiltonian around the $K_s$ point can be expressed as
\begin{equation}
H_{K_s}^{\alpha}(\k)=
\begin{pmatrix}
0  & \frac{(k_x+i \alpha k_y)^{M-1}}{m_{M-1}} &0  \;\\
\frac{(k_x-i \alpha k_y)^{M-1}}{m_{M-1}}  & 0 & t_{\textrm{eff}}\;\\
0 & t_{\textrm{eff}} & 0 \;
\end{pmatrix}\;,
\label{eq:hKs}
\end{equation}
where 
\begin{equation}
m_{M-1}\!=\!\frac{t_{\perp}^{M-2}}{(-1)^{M-2}(\hbar v_F)^{M-1}}\;,
\label{eq:mass}
\end{equation}
and $t_{\textrm{eff}}$ is the low-energy effectively coupling between the states of the $(M-1)$ layers and the flat bands from the interface TBG. In particular, if the stacking chirality is $+$,
  $t_{\textrm{eff}}$ represents the coupling between the Bloch states from the $B$  sublattice of the $(M-1)th$ layer and \textit{one of the two flat bands with $A$ sublattice polarization} contributed by the interface TBG. If the stacking chirality is $-$, then  $t_{\textrm{eff}}$ represents the coupling between the Bloch states from the $A$  sublattice of the $(M-1)th$ layer and \textit{one of the two flat bands with $B$ sublattice polarization} contributed by the interface TBG. Note that although $\k$ in Eq.~(\ref{eq:heff}) denotes the $\k$ point in the original primitive-cell BZ, while $\k$ in Eq.~(\ref{eq:hKs}) represents the $\k$ point in the moir\'e supercell BZ. This is because the form of the low-energy effective Hamiltonian (Eq.~(\ref{eq:heff})) is unchanged after the BZ folding.

Similarly one can write down the low-energy effective Hamiltonian around the $K_s'$ point, which consists of the two low-energy states from the $(N-1)$ layers above the TBG interface and \textit{one of the two flat bands} from the TBG interface. More explicitly, it can be expressed as
\begin{equation}
H_{K_s'}^{\alpha'}(\k)=
\begin{pmatrix}
0  & \frac{(k_x+i \alpha' k_y)^{N-1}}{m_{N-1}} & t_{\textrm{eff}}^{*} \;\\
\frac{(k_x-i\alpha' k_y)^{N-1}}{m_{N-1}} & 0 & 0\;\\
t_{\textrm{eff}} & 0 & 0 \;
\end{pmatrix}\;,
\label{eq:hKs2}
\end{equation}
where $\alpha'=\pm$ represents the stacking chirality of the $N$ layers, and $m_{N-1}$ is given by Eq.~(\ref{eq:mass}).

Both Eq.~(\ref{eq:hKs}) and Eq.~(\ref{eq:hKs2}) can be solved analytically. The eigenenergies of Eq.~(\ref{eq:hKs}) are expressed as:
\begin{align}
&\epsilon_{1\k}(M-1)\!=\!-\sqrt{m_{M-1}^2k^{2M-2}+\vert t_{\textrm{eff}}\vert^2}\;\nn
&\epsilon_{2\k}\!=\!0\;\nn
&\epsilon_{3\k}(M-1)\!=\!\sqrt{m_{M-1}^2k^{2M-2}+\vert t_{\textrm{eff}}\vert^2}\;.
\end{align}
The eigenenergies of Eq.~(\ref{eq:hKs2}) have exactly the same analytic expression but with $(M-1)$ replaced by $(N-1)$. The eigenstates of $H_{K_s}^{\alpha}(\k)$  are expressed as
\begin{align}
&\vert\psi_{1\k}^{M-1}\rangle=\Big(\,\frac{(k_x+i\alpha k_y)^{M-1}}{\sqrt{2}g_{M-1}(\vert\k\vert)}\, ,\,-\frac{1}{\sqrt{2}}\, ,\,\frac{\widetilde{t}_{\textrm{eff}}}{\sqrt{2}g_{M-1}(\vert\k\vert)}\,\Big)^{T}\;\nn
&\vert\psi_{2\k}^{M-1}\rangle=\Big(\,\frac{-\widetilde{t}_{\textrm{eff}}\,e^{i\alpha\theta_{\k}(M-1)}}{g_{M-1}(\k)}\,  , \,0\, ,\,\frac{\vert\k\vert^{(M-1)}}{g_{M-1}(\k)}\,\Big)^{T}\;\nn
&\vert\psi_{3\k}^{M-1}\rangle=\Big(\,\frac{(k_x+i\alpha k_y)^{M-1}}{\sqrt{2}g_{M-1}(\vert\k\vert)}\, ,\,\frac{1}{\sqrt{2}}\, ,\,\frac{\widetilde{t}_{\textrm{eff}}}{\sqrt{2}g_{M-1}(\vert\k\vert)}\,\Big)^{T}
\label{eq:psiM}
\end{align}
where $g_{M-1}(\vert\k\vert)\!=\!\sqrt{\widetilde{t}_{\textrm{eff}}^2+\vert\k\vert^{2M-2}}$,  $\widetilde{t}_{\textrm{eff}}\!=\!t_{\textrm{eff}}/m_{M-1}$, and $\theta_{\k}\!=\!\arctan(k_y/k_x)$.
The eigenstates of $H_{K_s'}^{\alpha'}(\k)$ are expressed as
\begin{align}
&\vert\psi_{1\k}^{N-1}\rangle=\Big(\,-\frac{1}{\sqrt{2}}\, ,\,\frac{(k_x-i\alpha' k_y)^{N-1}}{\sqrt{2}\,g_{N-1}(\vert\k\vert)}\, , \,\frac{\widetilde{t}_{\textrm{eff}}}{\sqrt{2}\,g_{N-1}(\vert\k\vert)}\,\Big)^{T}\;\nn
&\vert\psi_{2\k}^{N-1}\rangle=\Big(\,0\,  , \,-\frac{\widetilde{t}_{\textrm{eff}}\,e^{-i\alpha'\theta_{\k}(N-1)}}{g_{N-1}(\k)}\, ,\,\frac{\vert\k\vert^{(N-1)}}{g_{N-1}(\k)}\,\Big)^{T}\;\nn
&\vert\psi_{3\k}^{N-1}\rangle=\Big(\,\frac{1}{\sqrt{2}}\, ,\,\frac{(k_x-i\alpha' k_y)^{N-1}}{\sqrt{2}\,g_{N-1}(\vert\k\vert)}\, , \,\frac{\widetilde{t}_{\textrm{eff}}}{\sqrt{2}\,g_{N-1}(\vert\k\vert)}\,\Big)^{T}\;.
\label{eq:psiN}
\end{align}
Given that $(k_x\pm\alpha k_y)^{M-1}$ can be rewritten as $\vert\k\vert^{M-1}e^{i\pm\alpha\theta_{\k}(M-1)}$, it is straightforward to calculate the Berry connections of the valence states and the conduction states. For the states around the $K_s$ point (Eq.~(\ref{eq:psiM}), it turns out that
\begin{align}
A_{1\theta_{\k}}^{M-1}
= i\langle\psi_{1\k}^{M-1}\vert\partial_{\theta_\k}\psi_{1\k}^{M-1}\rangle
=-\frac{\alpha(M-1)\,\vert \k\vert^{2M-2}}{2\widetilde{t}_{\textrm{eff}}^2+ 2\vert\k\vert^{2M-2}},
\label{eq:berry-connection-1}
\end{align}
and $A_{3\theta_{\k}}^{M-1}\!=\!A_{1\theta_{\k}}^{M-1}$. For the states around the $K_s'$ point, we have
\begin{align}
A_{1\theta_{\k}}^{N-1}=i\langle\psi_{1\k}^{N-1}\vert\partial_{\theta_\k}\psi_{1\k}^{N-1}\rangle =\frac{\alpha'(N-1)\,\vert \k\vert^{2N-2}}{2\widetilde{t}_{\textrm{eff}}^2+ 2\vert\k\vert^{2N-2}},
\label{eq:berry-connection-2}
\end{align}
and $A_{3\theta_{\k}}^{N-1}\!=\!A_{1\theta_{\k}}^{N-1}$.
It is interesting to note that the valence and conduction states $\vert\psi_{1\k}\rangle$ and $\vert\psi_{3\k}\rangle$ have the same Berry connections, therefore they have the same Berry phase by virtue of the chiral symmetry of Eq.~(\ref{eq:hKs}) and Eq.~(\ref{eq:hKs2}). Taking the limit $\widetilde{t}_{\textrm{eff}}\!\to\!0$, from Eq.~(\ref{eq:berry-connection-1}-\ref{eq:berry-connection-2}) it follows that $A_{1\theta_{\k}}^{M-1}\!=\!A_{3\theta_{\k}}^{M-1}\!\to\!-\alpha(M-1)/2$, and $A_{1\theta_{\k}}^{N-1}\!=\!A_{3\theta_{\k}}^{N-1}\!\to\!\alpha'(N-1)/2$. Therefore, around $K_s$ ($K_s'$) point, the sum of the Berry fluxes of the conduction and the valence bands equals to $-\alpha(M-1)$ ($\alpha' (N-1)$). Then the total Chern number of the conduction and the valence bands equal to $-\alpha(M-1)+\alpha'(N-1)$. The Chern number of the two flat bands ($\vert\psi_{2\k}\rangle$) must cancel the total Chern number of the valence and conduction bands ($\vert\psi_{1\k}\rangle$ and $\vert\psi_{3\k}\rangle$), it follows that the total Chern number of the two flat bands equals to $\alpha(M-1)-\alpha'(N-1)$ for the $K$ valley. The total Chern number of the two flat bands of the $K'$ valley is just opposite to that of the $K$ valley, thus Eq.~(\ref{eq:chern-number}) has been proved.

\section{Calculating the local charge current using the continuum model}
\label{append:current}

The current density operator at a position $\mathbf{r}$ is expressed as
\begin{equation}
\hat{\mathbf{J}}(\mathbf{r})=e\hat{\rho}(\mathbf{r})\,\hat{\mathbf{v}}\;,
\label{eq:current-density}
\end{equation}
where $\hat{\rho}(\mathbf{r})$ is the local density operator at $\mathbf{r}$, and the velocity operator $\hat{\mathbf{v}}$ satisfies the Sch\"{o}rdinger equation $-i\hbar\hat{\mathbf{v}}\!=[H(\mathbf{r}),\mathbf{r}]$. The Hamiltonian at $\mathbf{r}$ is given by Eq.~(\ref{eq:HMN}), with $\mathbf{k}\!=\!-i\partial_{\mathbf{r}}$. Then it is straightforward to calculate the velocity operator by performing the commutator of $H(\mathbf{r})$ and $\mathbf{r}$.  
The expectation value of $\hat{\mathbf{J}}(\mathbf{r})$ then equals to
\begin{equation}
\langle \hat{\mathbf{J}}(\mathbf{r})\rangle
=e\sum_{s\mathbf{G} \k}\langle s\mathbf{G},\k\vert\,\hat{\rho}\, \hat{\rho}(\mathbf{r})\,\hat{\mathbf{v}}\,\vert s\mathbf{G},\k \rangle\;,
\end{equation}
where $\hat{\rho}\!=\!\sum_{n\k}\vert\psi_{n\k}\rangle\langle \psi_{n\k}\vert \theta(\mu-\epsilon_{n\k})$ is the density matrix at zero temperature with the chemical potential $\mu$, with $\vert\psi_{n\k}\rangle$ and $\epsilon_{n\k}$ being the $n$th eigenstate and eigenenergy of the Hamiltonian Eq.~(\ref{eq:HMN}) at the $\k$ point in the moir\'e BZ. $\vert s\mathbf{G},\k \rangle$ is the plane-wave function, where $\mathbf{G}$ represents a reciprocal lattice vector of the moir\'e cell, and $s$ is the index for the layer and sublattice degrees of freedom. To be more explicit, in the plane-wave basis, $\hat{\rho}$, $\rho(\mathbf{r})$ and $\hat{v}$ are expressed as
\begin{align}
&\langle s\mathbf{G},\k\vert\hat{\rho}\vert s'\mathbf{G}',\k\rangle
=\sum_{n\k}C_{s\mathbf{G},n}(\k)C_{s'\mathbf{G}',n}^{*}(\k)\theta(\mu-\epsilon_{n\k})\;\\
&\langle s\mathbf{G},\k\vert\hat{\rho}(\mathbf{r})\vert s'\mathbf{G}',\k\rangle=\frac{1}{V}\delta_{s s'} e^{-i\mathbf{G}\!\cdot\!\mathbf{r}}
e^{i\mathbf{G}'\!\cdot\!\mathbf{r}}\;\\
&\langle s\mathbf{G},\k\vert \hat{\mathbf{v}} \vert s'\mathbf{G}',\k\rangle
=\langle s\mathbf{G},\k\vert \frac{\partial H_{\k}}{\hbar \partial \k}
\vert s'\mathbf{G}',\k\rangle\;,
\end{align}
where the $C_{s\mathbf{G},n}(\k)$ is the plane-wave coefficient of the eigenstate $\vert\psi_{n\k}\rangle$, i.e., $\vert\psi_{n\k}\rangle=\sum_{s\mathbf{G}}C_{s\mathbf{G},n}(\k)\vert s\mathbf{G},\k\rangle$. $V$ is the total volume of the system,  and $H_{\mathbf{k}}=e^{-i\k\cdot\mathbf{r}}He^{i\k\cdot\mathbf{r}}$, where the Hamiltonian $H$ is given by Eq.~(\ref{eq:HMN}).   Given the current density distribution $\mathbf{J}(\mathbf{r})$, the magnetic field $B(\mathbf{r})$ can be solved using the Amp\`ere's law, $\nabla\times \mathbf{B}(\mathbf{r}) =\mu_0 \mathbf{J}(\mathbf{r})$, where $\mu_0$ is the magnetic permeability of the vacuum.

\bibliography{tmg}
\end{document}